\documentclass[aps,prl,twocolumn,groupedaddress]{revtex4-2}

\usepackage{amsmath}
\usepackage{amssymb}
\usepackage{amsfonts}
\usepackage[dvipsnames]{xcolor}
\usepackage{hyperref}
\hypersetup{
	pdftex,
	pdfstartview=FitH,
	bookmarksnumbered=true,
	pagebackref,
	colorlinks,
	breaklinks,
	urlcolor=Maroon,
	linkcolor=Maroon,
	citecolor=Bittersweet
} 
\usepackage{graphicx}
\DeclareGraphicsExtensions{.pdf,.png,.jpg,.mps}
\usepackage{tikz}
\usetikzlibrary{calc,matrix,decorations.pathmorphing,decorations.markings,arrows,positioning,intersections,mindmap,backgrounds}
\usepackage{booktabs}
\usepackage{physics}
\usepackage{diagbox}
\usepackage{cleveref}

\usepackage{lmodern} 
\usepackage{scalerel,accents,trimclip}
\usepackage{CJK}

\newcommand{\zb}{{\bar{z}}}
\newcommand{\vbar}{{\bar{v}}}

\newcommand{\Op}[1]{\mathcal{O}_{#1}}
\newcommand{\Jp}[1]{\mathcal{J}_{#1}}

\newcommand{\str}{{\rm str}}
\newcommand{\sdet}{{\rm sdet}}

\newcommand{\newhat}{\scalebox{2}[1.05]{\trimbox{0pt 1.15ex}{\textasciicircum}}}
\newcommand{\mhat}[1]{\accentset{\newhat}{#1}}
\newcommand{\newhatt}{\scalebox{1}[1.05]{\trimbox{0pt 1.0ex}{\textasciicircum}}}
\newcommand{\shat}[1]{\accentset{\newhatt}{#1}}
\newcommand{\sumint}{\scalebox{1.1}{$\displaystyle\int$}\hspace{-4mm}\scalebox{0.9}{$\displaystyle\sum$}}

\newcommand{\sectionsep}{\vspace{0.5em}}

\usepackage{graphicx}
\usepackage{bbold}

\begin{document}


	
	\begin{CJK*}{UTF8}{gbsn}

	\title{All Five-point Kaluza-Klein Correlators and Hidden 8d Symmetry in $\rm AdS_5\times S^3$}


	\author{Zhongjie Huang (黄中杰)}
	\email{zjhuang@zju.edu.cn}

	\author{Bo Wang (王波)}
	\email{b\_w@zju.edu.cn}
	\author{Ellis Ye Yuan (袁野)}
	\email{eyyuan@zju.edu.cn}
	\author{Jiarong Zhang (张家荣)}
	\email{steven\_rong@zju.edu.cn}
	\affiliation{Zhejiang Institute of Modern Physics, School of Physics, Zhejiang University, \\Hangzhou, Zhejiang 310058, China }
	\affiliation{Joint Center for Quanta-to-Cosmos Physics, Zhejiang University,
		\\Hangzhou, Zhejiang 310058, China}

	\date{\today}

	\begin{abstract}
		We systematically compute five-point correlators of chiral primary operators with arbitrary Kaluza--Klein charges at tree-level in $\mathrm{AdS}_5\times\mathrm{S}^3$, and obtain a unified formula. This result serves as the first concrete confirmation for the existence of the hidden eight-dimensional symmetries at the level of five points.
	\end{abstract}


	\maketitle
	
	\end{CJK*}


	\noindent{\bf Introduction and summary.} Correlators of chiral primary operators (CPOs) with various Kaluza--Klein (KK) charges have been one of the main themes in the study of holographic correlators. They help solve degeneracy in the OPE, and so serve as crucial ingredients in fully understanding the spectrum and interactions. For this reason, the remarkable discovery of a unified formula for arbitrary KK four-point tree-level correlators in $\mathcal{N}=4$ SYM in the supergravity limit \cite{Rastelli:2016nze,Rastelli:2017udc} has triggered broad advances in perturbative gravity scattering in AdS at higher precision \cite{Alday:2017xua,Aprile:2017bgs,Aprile:2017xsp,Alday:2017vkk,Aprile:2017qoy,Aprile:2018efk,Caron-Huot:2018kta,Goncalves:2019znr,Aprile:2019rep,Aprile:2020luw,Bissi:2020woe,Behan:2021pzk,Huang:2021xws,Drummond:2022dxw,Goncalves:2023oyx,Huang:2024rxr}. Similar developments have also taken place in other holographic models \cite{Giusto:2018ovt,Rastelli:2019gtj,Giusto:2019pxc,Giusto:2020neo,Alday:2020dtb,Alday:2020lbp,Alday:2021odx,Abl:2021mxo,Rigatos:2024yjo}.
	
	Recently there have been growing interests in holographic correlators at higher points \cite{Goncalves:2019znr,Alday:2022lkk,Goncalves:2023oyx,Alday:2023kfm,Cao:2023cwa,Cao:2024bky}. Besides the aim of exploring particle dynamics in AdS, these correlators can potentially uncover new data of the spectrum that were hardly reachable at four points. With this consideration, it will be ideal that unified formulas for higher-point Kazula--Klein correlators are available.
	
	In this paper we attack this problem at five points, in the gluon sector of a holographic model in $\mathrm{AdS}_5\times\mathrm{S}^3$. This model can be constructed, e.g., by inserting probe D7 branes in $\mathcal{N}=4$ SYM \cite{Karch:2002sh}\footnote{Alternatively this model can also be realized by probing an F-theory D7-brane singularity using a stack of D3-branes \cite{Fayyazuddin:1998fb,Aharony:1998xz}. }, preserving half amount of supersymmetries. We first bootstrap correlators of five CPOs in Mellin space. This is carried out for a whole class of correlators with fixed extremality at a time. By working out the lowest few extremality classes, we manage to find a closed formula \eqref{eq:M5} for the Mellin amplitudes of all KK modes. In position space, this can be expressed as a single generating function for all five-point correlators (see \eqref{eq:resultposition} for the explicit result)
	\begin{align*}
	    \mathcal{G}_5(x^2_{ij};v_{ij},\vbar_{ij}) =  \mathcal{G}_{5}^{\text{free}} +  \sum_{a} R_a(x^2_{ij};v_{ij},\vbar_{ij})  \mathcal{H}_a(X^2_{ij}),
	\end{align*}
	with $X^2_{ij} \equiv x^2_{ij}-t_{ij} =  x^2_{ij} - v_{ij}\vbar_{ij}$ to be some ``8d distance'', and $\mathcal{G}_{5}^{\text{free}}$ related to the free theory correlator obtained by Wick contractions. $R_a$ are polynomials of $x^2_{ij},v_{ij},\vbar_{ij}$, and automatically vanish once we set $x_{ij}^2 = t_{ij}$. To obtain the correlator with specific KK configuration, one Taylor expands $\mathcal{G}_5$ around $t_{ij}=0$ and reads off the coefficient $G_{p_1p_2p_3p_4p_5}$ with suitable exponents of $t_{ij}$'s. This result can be compared with the known four-point formula \cite{Alday:2021odx}
	\begin{align*}
	    \mathcal{G}_4(x^2_{ij};v_{ij},\vbar_{ij}) = \mathcal{G}_4^{\text{free}} + R(x^2_{ij};v_{ij}) \, \mathcal{H}(X^2_{ij})\, ,
	\end{align*}
	where $R$ follows the same properties as above and is dictated by the superconformal Ward identity \cite{Nirschl:2004pa}.

	\sectionsep\noindent{\bf Preliminaries.} The CPOs we focus on are
	\begin{align*}
        \mathcal{O}_{p}^I(x;v,\!\vbar)\equiv\mathcal{O}_p^{I;\alpha_1\ldots\alpha_p,\bar{\alpha}_1\ldots\bar{\alpha}_{p\!-\!2}}(x)\,v_{\alpha_1}\!\cdots v_{\alpha_p}\bar{v}_{\bar{\alpha}_1}\!\cdots\bar{v}_{\bar{\alpha}_{p\!-\!2}}.
    \end{align*}
	They correspond to supergluons ($p=2$) and their KK descendants ($p>2$) in AdS \footnote{In free field theory, these CPOs can be constructed from the scalars $q$, $\bar{q}$ in fundamental hypermultiplets and $\phi$ in the adjoint \cite{Huang:2023ppy}. Schematically we have $\mathcal{O}_p^{I} \sim \bar{q}\phi^{p-2}q$ .}. They have protected dimension $\Delta=p$, and come as the adjoint representation of a (bulk) gauge group $G_F$ \footnote{In the construction with F-theory singularity, $G_F$ can be $U(1)$, $SU(2)$, $SU(3)$, $SO(8)$, $E_6$, $E_7$, $E_8$, depending on the type of the singularity. In the D3-D7 construction, $G_F=SU(N_F)$. }, indicated by the index $I$.  These operators transform under the spin-$\frac{p}{2}$ and spin-$\frac{p-2}{2}$ representations of the R-symmetry group ${\rm SU(2)_R}$ and another flavor group ${\rm SU(2)_L}$. We use two-component bosonic spinors $v,\vbar$ to contract all ${\rm SU(2)_{R,L}}$ indices, so that correlators become functions on the scalar products
	\begin{gather*}
	    x^2_{ij}\! =\! (x_i\!-\!x_j)^2,\quad v_{ij}\! =\! \epsilon_{\alpha\beta}v_i^{\alpha}v_j^{\beta},\quad \vbar_{ij}\! =\! \epsilon_{\bar{\alpha}\bar{\beta}}\vbar_i^{\bar{\alpha}}\vbar_j^{\bar{\beta}}\,.
	\end{gather*}
	In the theory, there are also supergraviton excitations. They have weaker coupling to supergluons and are thus suppressed at the tree-level gluon scattering \cite{Kruczenski:2003be,Aharony:2007dj}. Since all CPOs are in the adjoint of $G_F$, we decompose correlators onto traces of generators $\tr[ijk\cdots]\equiv\tr(T^{I_i}T^{I_j}T^{I_k}\cdots)$, as commonly done in scattering amplitudes. At tree level we thus have
    \begin{align}\label{eq:colorDecomposition}
        \langle\mathcal{O}_1\mathcal{O}_2\cdots\mathcal{O}_n\rangle=\!\!\sum_{\sigma\in \mathcal{S}_{n}/\mathrm{Z}_n}\!\!\tr[\sigma]\,G[\sigma]+\cdots.
    \end{align}
    Here $G[\sigma]$ is called a partial correlator, endowed with a planar ordering $\sigma$ from $\tr[\sigma]$. ``$\cdots$'' denotes higher-trace terms that are in principle determined by the single-trace $G[\sigma]$ \footnote{Because graviton exchanges are suppressed, these higher-trace terms are purely effects of color decomposition. Hence they can be determined group theoretically from the single-trace ones \cite{Dixon:1996wi}, and it is not necessary to compute them separately. Because we only focus on the single-trace contributions, our computation applies universally to any $G_F$.}. Without loss of generality, below we only consider the partial correlator with canonical ordering
    \begin{align*}
        G_n(x^2_{ij};v_{ij},\vbar_{ij})\ \equiv\ G[12\cdots n]\,.
    \end{align*}

    \sectionsep\noindent{\bf Mellin amplitude and its factorization.}  Our main computation is performed in Mellin space. Mellin amplitude $M$ of a scalar correlator $G$ is defined as \cite{Mack:2009mi,Penedones:2010ue}
     \begin{equation}\label{eq:scalarmellin}
        G = \int  [\dd \gamma] M(\gamma_{ij})\prod_{i<j}  \frac{\Gamma(\gamma_{ij})}{( x^2_{ij} )^{\gamma_{ij}}}   \, ,
    \end{equation}
    subjecting to constraints
    \begin{align}\label{eq:gammaeq}
        \sum_i \gamma_{ij} = 0, \qquad \gamma_{ij}=\gamma_{ji}, \qquad \gamma_{ii} = -{p_i}\, ,
    \end{align}
    {$p_i$ being the dimension of the $i$-th operator.} The integration is on any set of independent $\gamma$'s
    \begin{align*}
        [\dd \gamma] = \prod\frac{\dd \gamma_{ij}}{2\pi i}\,.
    \end{align*}
    A virtue of the Mellin amplitude is that, its poles relate to exchange of operators in the corresponding channel, and the corresponding residues factorize into two sub-amplitudes \cite{Fitzpatrick:2011ia,Goncalves:2014rfa}. Explicitly, for a channel that separates operators $1,\ldots,k$ and $k+1,\ldots,n$, with a twist $p$ operator exchange, the Mellin amplitude behaves as \footnote{We temporarily omit all the structures with $G_F$, ${\rm SU(2)_R}$ and ${\rm SU(2)_L}$. These dependence will be restored later this section.}
    \begin{align}\label{eq:MellinPole}
        M(\gamma_{ij}) \sim \frac{Q_m}{\gamma_{LR}-p-2m}\,,\quad m=0,1,\dots\,,
    \end{align}
    for $\gamma_{LR} = \sum_{a=1}^k \sum_{b=k+1}^n \gamma_{ab}$ being around $p+2m$. The extra shift $2m$ is responsible for the conformal descendants of the operator. $Q_m$ can be constructed from Mellin amplitudes of $1,\ldots,k,p$ and $p,k+1,\ldots,n$. Especially, when the exchanged operator is a scalar,
    \begin{align*}
        Q_m = \frac{-2\Gamma(p)m!}{(p-1)_m} M_{L,m} \times M_{R,m}\,,
    \end{align*}
    where 
    \begin{align}\label{eq:ML}
        M_{L,m} = \sum_{\substack{i_{ab}\geq 0,\\ \Sigma i_{ab} = m}} M_{L}(\gamma_{ab}+i_{ab}) \prod_{1\leq a<b \leq k} \frac{(\gamma_{ab})_{i_{ab}}}{i_{ab}!}
    \end{align}
    and similarly for $M_{R,m}$. For the exchange of spinning operator, we refer readers to \cite{Goncalves:2014rfa}. 
    
    In the holographic context, as poles for multi-trace operators are already encoded in the $\Gamma$ factors in \eqref{eq:scalarmellin}, a tree-level $M$ itself only contains poles for single-particle operators \cite{Penedones:2010ue}. For each correlator, the exchange of single-particle operators is of finite number, which turns $M$ to be a rational function on $\gamma_{ij}$. These single-particle operators include all the KK CPOs $\mathcal{O}_p$, as well as their super-descendants $\mathcal{J}_p^\mu$ and $\mathcal{F}_p$, characterized by their quantum numbers as summarized in Table \ref{tab:components}. This requires us to consider three- and four-point amplitudes involving one external $\mathcal{J}$ or $\mathcal{F}$, which can be obtained from known formulas of $\langle \mathcal{O}\mathcal{O}\mathcal{O}\rangle$ and $\langle\mathcal{O}\mathcal{O}\mathcal{O}\mathcal{O}\rangle$ via analytic superspace \cite{Eden:2000qp,Bissi:2022wuh}. The strategy we utilize is to promote correlators of $\mathcal{O}$ into full super-correlators, from which we then extract desired component field correlators. {We provide these results in terms of Mathematica codes in the arXiv ancillary files.}
    \begin{table}[t]
        \centering
        \begin{tabular}{c|ccc}
            \toprule
            operator & $\mathcal{O}_p$ & $\mathcal{J}_p^\mu$ & $\mathcal{F}_p$ \\
            \midrule
            twist & $p$ & $p$ & $p+2$ \\
            \midrule
            \hspace{5pt}Lorentz spin\hspace{5pt} & $0$ & $1$ & $0$ \\
            \midrule
            $\mathrm{SU}(2)_{\rm R}$ spin & $\frac{p}{2}$ & $\frac{p}{2}-1$ & $\frac{p}{2}-2$ \\
            \midrule
            $\mathrm{SU}(2)_{\rm L}$ spin & \hspace{5pt}$\frac{p}{2}-1$\hspace{5pt} & \hspace{5pt}$\frac{p}{2}-1$\hspace{5pt} & \hspace{5pt}$\frac{p}{2}-1$\hspace{5pt} \\
            \midrule
            Rept.~of $G_F$ & {\bf adj.}  & {\bf adj.}  & {\bf adj.} \\
            \bottomrule
        \end{tabular}
        \caption{Single-particle operators in factorization.}
        \label{tab:components}
    \end{table}
    
    As in Table \ref{tab:components}, the exchanged operators usually contain flavor structures, making the above Mellin factorization story more complicated. For the gauge group $G_F$, since we are talking about partial amplitudes as in \eqref{eq:colorDecomposition}, the sub-amplitudes $M_L$ and $M_R$ are both partial amplitudes as well, with their planar orderings inherited from $M$. For $\mathrm{SU}(2)_{\rm R/L}$, in constructing residue at a pole, one should make sure how the $\mathrm{SU}(2)$ structures of the sub-amplitudes properly glue together. This can be most easily done in a polynomial basis, in which we set
    \begin{align}\label{eq:polybasis}
        v_i=\begin{pmatrix}
          1 \\ z_i
        \end{pmatrix}\,,
    \end{align}
    where now $v_{ij} = z_i - z_j$ and the $\mathrm{SU}(2)$ structures become polynomials on $z_i$. Especially, the two sub-amplitudes become polynomials on $\hat{z}$, the coordinate of the spinor for the exchanged operator
    \begin{equation*}
        L(\shat{z}) = \sum_{i=0}^n\, [L]_i\, \hat{z}^{\,i}, \qquad R(\hat{z}) = \sum_{i=0}^n\, [R]_i\, \hat{z}^{\,i}.
    \end{equation*}
    Here we denote $[P]_i$ as the coefficient of $\hat{z}^{\,i}$ in polynomial $P$. Gluing $L(\hat{z})$ and $R(\hat{z})$ means to integrate $\hat{z}$ out in an $\mathrm{SU}(2)$-invariant way, and the result is given by 
    \begin{equation}\label{eq:polyglue}
        \sum_{i=0}^n (-1)^{n-i}\, \frac{i!(n-i)!}{n!}\, [L]_i [R]_{n-i}\,.
    \end{equation}
    From the above result one can then recover the usual ${\rm SU(2)}$ structure written in terms of $v_{ij}$ and $\vbar_{ij}$ .
    
    Technical details in this section can be found in the supplemental material, including factorization formula of spin-1 operator, superspace computation of correlators with $\mathcal{J}$ and $\mathcal{F}$, and derivation of the gluing formula \eqref{eq:polyglue}.

\begin{figure}[h]
    \centering
    \includegraphics{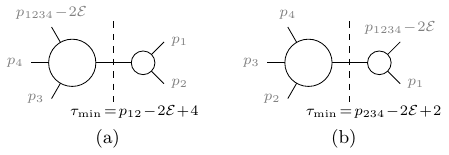}
    \caption{Minimal twist of exchanged $\mathcal{O}$.}
    \label{fig:minimalTwist}
\end{figure}

    \sectionsep\noindent{\bf Ansatz and bootstrap.} In order to simplify computations, it is practical to study correlators of a fixed extremality each time. Without loss of generality, in an $n$-point correlator we assume the $n$-th operator has the largest twist, then the extremality $\mathcal{E}$ is defined by \cite{DHoker:1999jke,Eden:2000gg}
    \begin{align*}
        p_n=p_1+p_2+\cdots+p_{n-1}-2\mathcal{E}.
    \end{align*}
    It is known that $\langle\mathcal{O}\mathcal{O}\mathcal{O}\rangle$ is non-vanishing only for $\mathcal{E}\geq1$ and $\langle\mathcal{O}\mathcal{O}\mathcal{O}\mathcal{O}\rangle$ only for $\mathcal{E}\geq2$ \cite{Alday:2021odx}. This fact strictly constraints the poles of Mellin amplitudes. At five points every factorization separates two operators (labeling them as $i$ and $j$) from the remaining three, and when we study the exchange of a twist $\tau$ operator, from \eqref{eq:MellinPole} we can see the corresponding pole locates at
    \begin{align*}
        \gamma_{ij}=\frac{p_i+p_j-\tau}{2}-m.
    \end{align*}
    At tree level, poles of $\gamma_{ij}$ in $M$ are bounded below by $\Gamma(\gamma_{ij})$ in \eqref{eq:scalarmellin}, and we need to work out the minimal allowed twist to set their upper bound. At fixed extremality, a single factorization channel falls into two types, depending on whether $5\in\{i,j\}$, see Figure \ref{fig:minimalTwist} (we abbreviate $p_{ij\cdots}\!\!\equiv\!\! p_i\!+\!p_j\!+\!\cdots$). For an $\mathcal{O}$ exchange, the set of allowed twists can be easily worked out by the above vanishing condition on the three- and four-point sub-amplitudes. The resulting minimal twist is shown in Figure \ref{fig:minimalTwist}. Since the $\mathcal{J}$ and $\mathcal{F}$ in the same multiplet cannot have lower twist, these are in fact the minimal twist for the entire channel. Therefore, for the two channels shown in Figure \ref{fig:minimalTwist} we expect the following set of poles
    \begin{subequations}\label{eq:singleChannel}
    \begin{align}
        \text{channel }(12):&\;\;\gamma_{12}-j,\qquad\quad\,\; j=1,2,\ldots,\mathcal{E}-2,\\
        \text{channel }(15):&\;\;\gamma_{15}-p_1+k,\quad k=1,2,\ldots,\mathcal{E}-1.\label{eq:channelb}
    \end{align}
    \end{subequations}
    Other channels follow similar patterns, depending on their types. Note that in \eqref{eq:channelb} (or Figure \ref{fig:minimalTwist}(b)) $k$ can in principle take even higher values. However, in practice one will observe that the residue $Q_m=0$ in the corresponding factorization, as the sub-amplitude \eqref{eq:ML} vanishes when its Mellin variables receive too high a shift (similar phenomenon has been observed in \cite{Goncalves:2023oyx}). So effectively the range of $k$ shrinks to the pattern shown above.

\begin{figure}[h]
    \centering
    \includegraphics{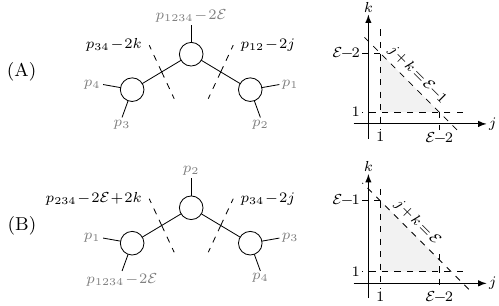}
    \caption{Range of twists in compatible factorizations.}
    \label{fig:twistRange}
\end{figure}

Five-point correlators also allow two consecutive OPEs, hence pair of compatible factorizations of $M_5$. They also fall into two types as shown in Figure \ref{fig:twistRange}, depending on whether the fifth operator locates in the three-point amplitude in the middle. In the diagrams in Figure \ref{fig:twistRange} we write out the twist of the exchanged operator following the convention in \eqref{eq:singleChannel}. In enumerating such compatible factorizations, we also need to make sure the two exchanged twists allow a non-vanishing three-point sub-amplitude in the middle. This yields an extra bound on the value of $j+k$: $j+k\leq \mathcal{E}-1$ for type (A) and $j+k\leq\mathcal{E}$ for type (B). Hence the values of $\{j,k\}$ are constrained within the shaded region in Figure \ref{fig:twistRange}.

{In addition, in the flat space limit where one sends $\gamma_{ij}\to \infty$, the Mellin amplitude reduces to ordinary scattering amplitude \cite{Penedones:2010ue}. This puts constraints on the power counting of the whole ansatz, requiring it to be at most $\order{\gamma^{-1}}$ \cite{Alday:2022lkk,Cao:2023cwa,Cao:2024bky}.} This rules out regular terms in $M_5$, and states that residues are at most linear in $\gamma$'s on simultaneous poles (corresponding to compatible factorizations), and are constants on single poles.

    
Now we are prepared to construct an ansatz for the five-point computation. First focus on the kinematic part. We target on the canonical color-ordered partial amplitude $M_5$, so it is convenient to choose $\{\gamma_{12},\gamma_{23},\gamma_{34},\gamma_{45},\gamma_{15}\}$ as independent Mellin variables, each corresponding to one physical channel of $M_5$. Following Figure \ref{fig:twistRange} we write down simultaneous poles \footnote{In the numerator it is not necessary to list $\gamma$'s that already appear in the denominator.}
\begin{subequations}
\begin{align}
    \text{(A)}:\;&\frac{\{1,\gamma_{23},\gamma_{45},\gamma_{15}\}}{(\gamma_{12}\!-\!j)(\gamma_{34}\!-\!k)},\quad\; j,k\geq 1,\,j\!+\!k\leq\mathcal{E}\!-\!1,\\
    \text{(B)}:\;&\frac{\{1,\gamma_{12},\gamma_{23},\gamma_{45}\}}{(\gamma_{34}\!-\!j)(\gamma_{15}\!-\!p_1\!+\!k)},\;\;\; j,k\geq 1,\,j\!+\!k\leq\mathcal{E},
\end{align}
\end{subequations}
and three other simultaneous poles of type (B). In addition we also write down single pole terms with numerical numerators, in correspondence to \eqref{eq:singleChannel}. Let us name the set of all these kinematic functions $\mathcal{K}$.

We then deal with $\mathrm{SU}(2)_{\rm R/L}$ structures. In principle, for each function in $\mathcal{K}$ we can work out the $v$ structure that is consistent with its factorization, but this is not computationally convenient. Instead, it is more convenient to perform the computation in polynomial basis \eqref{eq:polybasis}. Note that since the fifth operator has the largest twist (thus the largest amount of $v$), most $v$ contractions are of the form $v_{5i},\ i=1,\ldots,4$. So we work in the frame
\begin{align*}
    v_1\!=\!\begin{pmatrix}
          1 \\ 0
        \end{pmatrix}\!,\; 
    v_2\!=\!\begin{pmatrix}
          1 \\ 1
        \end{pmatrix}\!,\; 
    v_3\!=\!\begin{pmatrix}
          1 \\ z_1
        \end{pmatrix}\!,\; 
    v_4\!=\!\begin{pmatrix}
          1 \\ z_2
        \end{pmatrix}\!,\; 
    v_5\!=\!\begin{pmatrix}
          0 \\ 1
        \end{pmatrix}\!,
\end{align*}
where $v_{5i}=1$, and the $\mathrm{SU}(2)_{\rm R}$ structures directly become polynomials on $z_1$ and $z_2$ with total degree not higher than $\mathcal{E}-1$. The same method works for $\mathrm{SU}(2)_{\rm L}$ structures and gives $w_1$, $w_2$ polynomials with total degree not higher than $\mathcal{E}-3$. The set of monomials $z_1^m z_2^{n}w_1^i w_2^{j}$ ($m+n\leq\mathcal{E}-1$ and $i+j\leq\mathcal{E}-3$) then provides an (over-complete) basis for the flavor structures, and we denote it as $\mathcal{I}$.

With the above ingredients, we construct the ansatz for $M_5$ as a linear combination of
\begin{align*}
    \mathcal{K}\otimes\mathcal{I}
\end{align*}
with unknown coefficients which can depend on $p_i$. Based on this ansatz, we check all the five physical channels, and compare residue of the ansatz at each pole with the one worked out from factorization as described before. Note that in computing the residues from factorization one needs to make sure to include all possible contributions from $\mathcal{O}$, $\mathcal{J}$, $\mathcal{F}$ as well as their conformal descendants. Moreover, the residues on the simultaneous poles serve as consistency checks: there are two ways to factorize a simultaneous pole into three- and four-point amplitudes, and both ways should give the same residue. When all these comparisons are properly performed, we find that at each extremality the ansatz is solved completely.

    \sectionsep\noindent{\bf Generalized Mellin amplitude.}
    Using the above method, we first work out a few classes of correlators ($\mathcal{E}$=3,4,5,6, provided in the ancillary file), and observe a simple pattern in the residues on the poles. For example, for the pole at $\gamma_{12} = \mathcal{E}-2$ and $\gamma_{45}=p_4-2$, it is straightforward to see that 
    \begin{align*}
        M_{\{p_i\}} \supset &\ \frac{t_{15}^{\,p_1-\mathcal{E}+1}\,t_{25}^{\,p_2-\mathcal{E}+1}\,t_{35}^{\,p_3-2}\,t_{45}^{\,p_4-2}}{(p_1-\mathcal{E}+1)!\,(p_2-\mathcal{E}+1)!\,(p_3-2)!\,(p_4-3)!} \nonumber \\
        & \times  \frac{t_{12}^{\,\mathcal{E}-3}}{(\mathcal{E}-3)!} \frac{v_{12}^{\,2}\,v_{34}\,v_{45}\,v_{53}\,  \gamma_{23}}{(\gamma_{12}-\mathcal{E}+2)(\gamma_{45}-p_4+2)}\,,
    \end{align*}
    with $t_{ij} = v_{ij}\vbar_{ij}$. Such pattern suggests us to consider the $\rm AdS \times S$ Mellin formalism \cite{Aprile:2020luw}. In this formalism, apart from transforming $x_{ij}^2$ to $\gamma_{ij}$, there is also a (discrete) Mellin transformation from $t_{ij}$ to $n_{ij}$
    \begin{align*}
        H_{\{p_i\}}\!(x^2_{ij},t_{ij}) \!=& \sumint \widetilde{M}_{\{p_i\}}\!(\gamma_{ij},n_{ij}) \prod_{i<j} \frac{\Gamma(\gamma_{ij})}{\Gamma(n_{ij}\!+\!1)} \frac{t_{ij}^{\,n_{ij}}}{(x_{ij}^2)^{\gamma_{ij}}},
    \end{align*}
    where the sum is over all integers $n_{ij}$ satisfying 
    \begin{align*}
        \sum_{j} n_{ij} = 0\,, \quad n_{ij}=n_{ji}\,, \quad n_{ii} = -(p_i-2) \,.
    \end{align*}
    To go back to the usual Mellin amplitude one simply performs the sum over $n_{ij}$, and the $\Gamma$ functions in the denominator naturally set a cutoff on the range of $n_{ij}$ and give a finite sum. When $\widetilde{M}_{p_i}(\gamma_{ij},n_{ij})=\widetilde{M}(\rho_{ij})$ is only a function of the combinations $\rho_{ij} = \gamma_{ij} - n_{ij}$ but not $\{p_i\}$, one can further sum over all $p_i$ in the formula above. The resulting
    \begin{align*}
        \mathcal{H} = \sum_{p_i=0}^\infty H_{\{p_i\}} =&\ \int\  [\dd\rho]\, \widetilde{M}(\rho_{ij}) \prod_{i<j} \frac{\Gamma(\rho_{ij})}{(x_{ij}^2-t_{ij})^{\rho_{ij}}}
    \end{align*}
    is a generating function depending only on $X^2_{ij} = x^2_{ij}-t_{ij}$ \cite{Abl:2020dbx,Glew:2023wik}, manifesting the so-called hidden 8d symmetry in the literature. In this formalism, at four-point we have \cite{Drummond:2022dxd}
    \begin{align*}
        \widetilde{M}_{4} = \frac{1}{(\rho_{12}-1)(\rho_{14}-1)}\,,
    \end{align*}
    and the partial Mellin amplitude is given by 
    \begin{align*}
        M_{4} = \mhat{R}_{1234}  \circ  \widetilde{M}_{4} \, 
    \end{align*}
    (we leave the sum over $n_{ij}$ against $\prod_{i<j}\frac{(t_{ij})^{n_{ij}}}{(n_{ij})!}$ to be implicit). Here $\mhat{R}_{1234}$ is the Mellin version of the factor $R$ in the generating function $\mathcal{G}_4$, whose action is given by 
    \begin{align*}
        \mhat{R}_{1234} =  V_{1234}\, \shat{\gamma}_{13} \shat{\gamma}_{24} + V_{1342}\, \shat{\gamma}_{14} \shat{\gamma}_{23} + V_{1423}\,  \shat{\gamma}_{12} \shat{\gamma}_{34}\, ,
    \end{align*}
    $\shat{\gamma}_{ij}$ being a difference operator 
    \begin{align*}
        \shat{\gamma}_{ij} \circ F(\gamma_{ij},n_{ij}) = \gamma_{ij}\, F(\gamma_{ij}+1,n_{ij})\,,
    \end{align*}
    and $V_{i_1\cdots i_n}$ a shorthand for 
    \begin{align*}
        V_{i_1\cdots i_n} = v_{i_1i_2} v_{i_2i_3}\cdots v_{i_n i_1}\,.
    \end{align*}
    
    \sectionsep\noindent{\bf A unified formula for all KK modes.}
    With the $\rm AdS\times S$ Mellin transformation, at five points we find a unified result for all KK charge configurations
    \begin{align}\label{eq:M5}
        M_5 =&\ \mhat{R}^{(1)} \circ \widetilde{M}_{5}^{(1)} + \mhat{R}^{(2)} \circ \widetilde{M}_{5}^{(2)} + (\text{cyclic})\, ,
    \end{align}
    where
    \begin{align*}
        \widetilde{M}_{5}^{(1)} = -\frac{1}{5\,(\rho_{12}-1)(\rho_{23}-1)(\rho_{34}-1)}\,, \\
        \widetilde{M}_{5}^{(2)} = -\frac{2}{5\,(\rho_{12}-1)(\rho_{23}-1)(\rho_{45}-1)}\,,
    \end{align*}
    with
    \begin{align*}
        \mhat{R}^{(1)} \!=\! &\ \mhat{R}_{1234,5} \!+\! \mhat{R}_{2345,1} \!+\! \mhat{R}_{3451,2} \!+\! \mhat{R}_{4512,3} \!+\! \mhat{R}_{5123,4} \nonumber \\
        &+\!\left(\mathrm{v}_{13,5}\, \shat{n}_{13}\!+\!\mathrm{v}_{23,5}\,\shat{n}_{23}\!+\!\mathrm{v}_{14,5}\,\shat{n}_{14}\!+\!\mathrm{v}_{24,5}\,\shat{n}_{24}\right)\!\mhat{R}_{1234} \nonumber \\
        &+\!\left( \mathrm{v}_{13,4}\,\shat{n}_{13}\!+\!\mathrm{v}_{23,4}\,\shat{n}_{23}\!+\!\mathrm{v}_{53,4}\,\shat{n}_{53} \right)\! \mhat{R}_{1235} \nonumber \\
        &+\!\left( \mathrm{v}_{23,1}\,\shat{n}_{23}\!+\!\mathrm{v}_{24,1}\,\shat{n}_{24}\!+\!\mathrm{v}_{25,1}\,\shat{n}_{25} \right)\! \mhat{R}_{2345}\,, 
    \end{align*}
    and
    \begin{align*}
        \mhat{R}^{(2)} \!=\! &\ \mhat{R}_{4512,3} \!+\! \mhat{R}_{4513,2} \!+\! \mhat{R}_{4521,3} \!+\! \mhat{R}_{4523,1} \!+\! \mhat{R}_{4531,2}  \nonumber \\
        &+\! \mhat{R}_{4532,1} \!+\!\left( \mathrm{v}_{14,5}\,\shat{n}_{14}\!+\!\mathrm{v}_{24,5}\,\shat{n}_{24}\!+\!\mathrm{v}_{34,5}\,\shat{n}_{34} \right)\!\mhat{R}_{1234} \nonumber \\
        &+\!\left( \mathrm{v}_{51,4}\,\shat{n}_{51}\!+\!\mathrm{v}_{52,4}\,\shat{n}_{52}\!+\!\mathrm{v}_{53,4}\,\shat{n}_{53} \right)\! \mhat{R}_{1235}\,.
    \end{align*}
    Here we denote $\mathrm{v}_{ij,k} = v_{ik}v_{jk}v_{ij}^{-1}$ and
     \begin{align*}
        \shat{n}_{ij} \circ F(\gamma_{ij},n_{ij}) = n_{ij}\, F(\gamma_{ij},n_{ij}-1)\,.
    \end{align*}
    The five-label structure 
    \begin{align*}
        \mhat{R}_{1234,5} =&\ V_{12345}\, \shat{\gamma}_{14} \shat{\gamma}_{25}\shat{\gamma}_{35} + V_{12354}\, \shat{\gamma}_{34} \shat{\gamma}_{15}\shat{\gamma}_{25} \nonumber\\
        &+ V_{12534}\, \shat{\gamma}_{23} \shat{\gamma}_{15}\shat{\gamma}_{45} + V_{15234}\, \shat{\gamma}_{12} \shat{\gamma}_{35}\shat{\gamma}_{45}\, 
    \end{align*}
    transforms under the dihedral group of the first four labels, being invariant under cycling, and acquiring a minus sign under reflection. Note here in \eqref{eq:M5} the fifth operator is no longer special and the result enjoys cyclic symmetry. One can check that, by setting $p_i=2$ and performing the sum over $n_{ij}$ (which is equivalent to taking all $n_{ij}=0$), we reproduce the Mellin amplitude of $\langle 22222 \rangle$ in \cite{Alday:2022lkk}.
    
    From \eqref{eq:M5} we can go back to the position space by transforming both $\widetilde{\mathcal{M}}_{5}^{(1,2)}$ and $\mhat{R}^{(1,2)}$ back to functions of $x^2_{ij}$, $v_{ij}$ and $\vbar_{ij}$. For $\mhat{R}$ this simply amounts to replacing $\shat{\gamma}_{ij}\rightarrow x^2_{ij}$ and $\shat{n}_{ij}\rightarrow t_{ij}$, e.g.,
    \begin{align*}
        \mhat{R}_{1234} \rightarrow R_{1234} =&\ V_{1234}\, x^2_{13} x^2_{24} + V_{1342}\, x^2_{14} x^2_{23} \nonumber\\
        & +  V_{1423}\,  x^2_{12} x^2_{34} \, .
    \end{align*}
    For $\widetilde{\mathcal{M}}_{5}^{(1,2)}$ caution should be taken, because different terms in $\mhat{R}$ shifts the Mellin variables by different amounts, so that for each term the variables $\rho_{ij}$ in $\widetilde{\mathcal{M}}_{5}^{(1,2)}$ satisfy different constraints. Therefore, although the reduced amplitudes $\widetilde{\mathcal{M}}_{5}^{(1,2)}$ appear to be the same, the effective twists of them vary between different terms in $\mhat{R}$. Taking this into consideration, we get the generating function $\mathcal{G}_5$ for all KK modes (up to $\mathcal{G}_5^{\text{free}}$)
    \begin{widetext}
    \begin{align}\label{eq:resultposition}
        \mathcal{G}_5 -\mathcal{G}_5^{\text{free}} =&\  -\frac{1}{5X_{12}^2X_{23}^2X_{34}^2} \Big( R_{1234,5} \,D_{21124} + R_{2345,1} \,D_{31123} + R_{3451,2} \,D_{22123} + R_{4512,3} \,D_{21223} + R_{5123,4} \,D_{21133} \nonumber\\[1mm]
        &\qquad\qquad   + R_{1234} \big( \mathrm{v}_{13,5}\, t_{13}  \,D_{31222}+\mathrm{v}_{23,5}\,t_{23}\,D_{22222}+\mathrm{v}_{14,5}\,t_{14}\,D_{31132} +\mathrm{v}_{24,5}\,t_{24}\,D_{22132} \big) \nonumber \\[1mm]
        & \qquad\qquad  + R_{1235} \big( \mathrm{v}_{13,4}\,t_{13}\,D_{31213}+\mathrm{v}_{23,4}\,t_{23}\,D_{22213}+\mathrm{v}_{53,4}\,t_{53}\,D_{21214} \big) \nonumber \\[1mm]
        & \qquad\qquad  + R_{2345} \big( \mathrm{v}_{23,1}\,t_{23}\,D_{12223}+\mathrm{v}_{24,1}\,t_{24}\,D_{12133}+\mathrm{v}_{25,1}\,t_{25}\,D_{12124} \big) \Big) \nonumber \\
        & -\frac{2}{5X_{12}^2X_{23}^2X_{45}^2} \Big( R_{4512,3}\,D_{21322} + R_{4513,2}\,D_{22222} + R_{4521,3}\,D_{21322} + R_{4523,1}\,D_{31222} + R_{4531,2} \,D_{22222} \nonumber\\[1mm]
        &\qquad\qquad  + R_{4532,1}\,D_{31222} + R_{1234}\big( \mathrm{v}_{14,5}\,t_{14}\,D_{31231}+\mathrm{v}_{24,5}\,t_{24}\,D_{22231}+\mathrm{v}_{34,5}\,t_{34}\,D_{21331} \big) \nonumber \\[1mm]
        &\qquad\qquad + R_{1235} \big( \mathrm{v}_{51,4}\,t_{51}\,D_{31213}+\mathrm{v}_{52,4}\,t_{52}\,D_{22213}+\mathrm{v}_{53,4}\,t_{53}\,D_{21313} \big) \Big) + (\text{cyclic})\,.
    \end{align}
    \end{widetext}
    Here $D_{p_1p_2p_3p_4p_5}\equiv D_{p_1p_2p_3p_4p_5}(X^2_{ij})$ is the five-scalar contact Witten diagram in ${\rm AdS_5}$ with dimensions $\{p_i\}$ \cite{Penedones:2010ue}. These $D$-functions are normalized such that their Mellin amplitudes are simply 1. 
    
    Result \eqref{eq:resultposition} is a direct generalization of the four-point result, showing that the hidden 8d symmetry still exists for higher-point correlators at tree level. Unlike at four points, there are several $R_k$ structures in $\mathcal{G}_5$. Each $R_k$ individually manifests the chiral twist condition, i.e., they vanish upon placing all $x_i$ on a plane at $z_i$ and setting $v_i = \qty(\substack{1\\ z_i})$. In fact, they vanish under a more general configuration: if we take all $x_{i}$ to be light-like as $x_{i}^{\alpha\bar{\alpha}} = v_i^{\alpha}\vbar_i^{\bar{\alpha}}$, which is equivalent to setting $x^2_{ij}=t_{ij}$. This seems to be an analogy of Drukker-Plefka twist \cite{Drukker:2009sf} for $\mathcal{N}=2$ superconformal algebra, and we leave it for future exploration. 
    
    It is not surprising that there are more $R_a$ structures in the correlators, since the number of independent superinvariants increases. These structures, especially the five-label one $R_{1234,5}$, may serve as important ingredients in solving the superconformal Ward identity for five- and higher-point correlators. It would also be interesting to consider if there is a way to single out one particular $R_a$ structure when studying the correlator, to make the hidden 8d symmetry manifest in a simpler way. This may be achieved by taking certain superspace configurations, like considering the super-descendant of some $\Op{p}$.

    \sectionsep\noindent{\bf Outlook.} There are several future directions which are worthy of explorations. Our five-point result in principle encodes lots of new data for the operator spectrum and interactions. In particular, this should be the first place where one can make concrete observations on the properties of triple-trace operators, which by unitary cut are important for two-loop dynamics \cite{Bissi:2020wtv,Bissi:2020woe,Huang:2021xws,Drummond:2022dxw,Huang:2023oxf}. It is of immediate interest to learn how to properly extract these data from the five-point correlators and solve the related mixing problem.
    
    In this paper we extend the universal structure of CPO correlators from four to five points. It will be interesting to see how such structure further generalizes and whether it may help simplify computations at higher points.
    
    Moreover, our work leaves the possibility of better ways to organize the color-resummed full correlator. Relatedly, while in the current model we make use of color decomposition to simplify the computation, one has to face this problem when dealing with models like $\mathcal{N}=4$ super Yang--Mills. It is also interesting and important to find if similar structure occurs in this latter model.

    \sectionsep  
    \begin{acknowledgments}
    	\noindent{\bf Acknowledgments.}
    	The authors would like to thank Qu Cao for collaboration at the early stage of this work, and Hao Chen, Michele Santagata, Yichao Tang, Xinan Zhou for useful discussions, and Vasco Gon\c{c}alves, Konstantinos C.~Rigatos, Xinan Zhou for commenting on the first draft. The authors are supported by National Science Foundation of China under Grant No.~12175197 and Grand No.~12347103. EYY is also supported by National Science Foundation of China under Grant No.~11935013, and by the Fundamental Research Funds for the Chinese Central Universities under Grant No.~226-2022-00216.
    \end{acknowledgments}
    	
    \bibliography{refs}
    
    \widetext

    \begin{appendix}
    \begin{center}
        \textbf{\large Supplemental Material}
    \end{center}
    
    \section{Factorization of Mellin amplitude under spinning exchange}
    
    To conveniently take care of the spinning operator, we use a null polarization vector $z_\mu$ to contract its Lorentz index, i.e., $\mathcal{J}(x,z)\equiv z_\mu\mathcal{J}^\mu(x)$. The Mellin amplitude for a correlator involving one such spinning operator can then be defined by \cite{Goncalves:2014rfa} 
    \begin{align*}
        \langle \mathcal{J}(x_0,z) \mathcal{O}_1(x_1) \cdots \mathcal{O}_k(x_k)  \rangle = \sum_{a=1}^k -\frac{z\cdot x_{0a}}{x_{0a}^2} \int  [\dd \gamma] \gamma_{0a}M^a (\gamma_{ij})\prod_{i<j}( x^2_{ij} )^{-\gamma_{ij}} \Gamma(\gamma_{ij}) \, .
    \end{align*}
    where $x_{ij}^\mu\equiv x_i^\mu-x_j^\mu$, and the Mellin variables $\gamma$ satisfy the same set of linear relations as in the scalar case 
    \begin{align*}
        \sum_i \gamma_{ij} = 0,\qquad \gamma_{ij}=\gamma_{ji}, \qquad \gamma_{ii} = -\tau_i.
    \end{align*}
    Transversal property of the spinning polarization dictates that the component amplitudes satisfy the constraint
    \begin{align}\label{eq:transversality}
        \sum_{a=1}^k \gamma_{0a}M^a = 0.
    \end{align}
    For a spin-1 operator $\Jp{p}$ exchange in a scalar correlator, the residue at the twist of  $\Jp{p}$ is
      \begin{align}\label{eq:spin1fac}
        \mathcal{Q}_m =\frac{(p+1)\Gamma(p)m!}{(p)_m} \sum_{a=1}^k \sum_{b=k+1}^n M^a_{L,m} M^b_{R,m}\left[ \gamma_{ab} - \frac{p-1}{m(p-2)}\sum_{a'=1}^k \gamma_{aa'}\! \sum_{b'=k+1}^n \gamma_{bb'} \right],
      \end{align}
      where $M^a_{L,m}$ and $M^a_{R,m}$ is defined in the same way as $M_{L,m}$ and $M_{R,m}$
      \begin{align*}
        M^a_{L,m} = \sum_{\substack{i_{ab}\geq 0,\\ \Sigma i_{ab} = m}} M^a_{L}(\gamma_{ab}+i_{ab}) \prod_{1\leq a<b \leq k} \frac{(\gamma_{ab})_{i_{ab}}}{i_{ab}!},\qquad
        M^a_{R,m} = \sum_{\substack{i_{ab}\geq 0,\\ \Sigma i_{ab} = m}} M^a_{R}(\gamma_{ab}+i_{ab}) \prod_{k+1\leq a<b \leq n} \frac{(\gamma_{ab})_{i_{ab}}}{i_{ab}!}.
    \end{align*}
      Note that for $m=0$ the second term in the parenthesis vanishes.

    \section{Correlators with super-descendant operator}
    
    In preparation for the computation at five points, we should work out three- and four-point correlators involving one super-descendant field, i.e., $\langle \mathcal{J}\mathcal{O}\mathcal{O}\rangle$, $\langle\mathcal{J}\mathcal{O}\mathcal{O}\mathcal{O}\rangle$, $\langle \mathcal{F}\mathcal{O}\mathcal{O}\rangle$ and $\langle\mathcal{F}\mathcal{O}\mathcal{O}\mathcal{O}\rangle$. These data are needed in studying the factorization of $\langle\mathcal{O}\mathcal{O}\mathcal{O}\mathcal{O}\mathcal{O}\rangle$ when the exchanged operator is $\mathcal{J}$ or $\mathcal{F}$. The basic strategy to achieve this is to uplift known correlators of super primaries $\langle\mathcal{O}\mathcal{O}\mathcal{O}\rangle$ and $\langle\mathcal{O}\mathcal{O}\mathcal{O}\mathcal{O}\rangle$ to analytic superspace following \cite{Eden:2000qp}, and then read off the components with proper combinations of superspace Grassmann variables.

    In analytic superspace, the superspace coordinates is represented by a supermatrix
    \begin{align*}
        \mathbf{X} = \begin{pmatrix}
          x_{\alpha\dot{\alpha}} & \lambda_\alpha\\
          \pi_{\dot{\alpha}} & y
        \end{pmatrix},
    \end{align*}
    where $x_{\alpha\dot{\alpha}} = x_\mu \sigma^\mu_{\alpha\dot{\alpha}}$ is a $2\times 2$ matrix with components (in Euclidean signature)
    \begin{align*}
        x_{\alpha\dot{\alpha}} = \begin{pmatrix}
          x^0+ix^3 &\  ix^1+x^2 \\ ix^1-x^2 &\  x^0-ix^3
        \end{pmatrix},
    \end{align*}
    $\lambda_\alpha$ and $\pi_{\dot{\alpha}}$ are Grassmann variables, and $y$ is a number encoding the R-symmetry polarization, relate to the $\mathrm{SU}(2)_{\rm R}$ spinor by $v=\qty(\substack{1\\ y})$.

    One important superspace covariant, the superpropagator, is given by the superdeterminant of $\mathbf{X}^{-1}_{ij}= (\mathbf{X}_i-\mathbf{X}_j)^{-1}$ 
    \begin{align*}
        \hat{g}_{ij}\equiv\sdet \mathbf{X}_{ij}^{-1} = \frac{y_{ij}- \pi_{ij} x_{ij}^{-1}\lambda_{ij}}{x_{ij}^2} .
    \end{align*}
    Setting $\lambda=\pi=0$ gives us the familiar combination $y_{ij}/x_{ij}^2\equiv v_{ij}/x_{ij}^2\equiv g_{ij}$. For four-point correlators, we have three superinvariants (different with that in \cite{Eden:2000qp} under certain permutations) 
    \begin{align*}
        \hat{A}=&\ \sdet(\mathbf{X}^{\vphantom{1}}_{12}\mathbf{X}^{-1}_{24}\mathbf{X}^{\vphantom{1}}_{43}\mathbf{X}^{-1}_{31}),\qquad \hat{B}=\ \sdet(\mathbf{X}^{\vphantom{1}}_{14}\mathbf{X}^{-1}_{42}\mathbf{X}^{\vphantom{1}}_{23}\mathbf{X}^{-1}_{31}),\qquad \hat{C}=\ \str(\mathbf{X}^{\vphantom{1}}_{12}\mathbf{X}^{-1}_{24}\mathbf{X}^{\vphantom{1}}_{43}\mathbf{X}^{-1}_{31}).
    \end{align*}
    With these we can construct the supersymmetric version of cross-ratios $U$, $V$ and $\alpha$
    \begin{align*}
        \hat{U} &= \frac{1-\hat{B}-\hat{C}}{1-\hat{A}-\hat{B}}\hat{A},\qquad \hat{V} = \frac{\hat{C}-\hat{A}}{1-\hat{A}-\hat{B}}\hat{B},\qquad \hat{\alpha} = \frac{1-\hat{B}-\hat{C}}{1-\hat{A}-\hat{B}}.
    \end{align*}
    The uplift of correlators of the super primaries $\mathcal{O}$ to the super-correlator of the corresponding superfield $\hat{\mathcal{O}}$ can then be performed by directly replacing $g$, $U$, $V$ and $\alpha$ to the above supersymmetric version. Especially, we do this by first writing correlators in terms of their Mellin amplitudes and do the replacement for the Mellin transformation.
     
    When extracting correlators with one super-descendant operator from the super-correlator, one needs to pay attention that at the same Grassmann level there can be contributions from various components. For example, we expect $\mathcal{J}^\mu$ to appear at the level $\lambda\pi$, but at this level the whole contribution to $\hat{\mathcal{O}}$ is 
    \begin{align}
        \hat{\mathcal{O}}\supset\lambda_\alpha\pi_{\dot\alpha}\,\bar{\sigma}_\mu^{\dot{\alpha}\alpha}\qty(C_1\mathcal{J}^\mu+C_2\partial^\mu\partial_y\mathcal{O}),
    \end{align}
    where $C_1$ and $C_2$ are some numerical coefficients. Like in \cite{Bissi:2022wuh}, we can cleanly extract $\mathcal{J}^\mu$ by defining a differential operator $(\mathcal{D}_{\mathcal{J}})^\mu$ such that $\mathcal{J}^\mu=(\mathcal{D}_{\mathcal{J}})^\mu\hat{\mathcal{O}}$ (assuming that we set all Grassmann variables to zero after its action). Such differential operator is given by 
    \begin{align*}
        (\mathcal{D}_{\mathcal{J}})^\mu = \frac{\sqrt{2}}{\sqrt{(p+1)(p-1)}}\left( \frac{1}{2}\sigma^{\mu}_{\alpha\dot\alpha}\partial^{\dot{\alpha}} \partial^{\alpha} + \frac{1}{p}\, \partial^\mu \partial_y \right),
    \end{align*}
    where $\partial^\alpha=\partial/\partial\lambda_\alpha$ and $\partial^{\dot\alpha}=\partial/\partial\pi_{\dot\alpha}$. The normalization of $(\mathcal{D}_{\mathcal{J}})^\mu$ matches the normalization of spinning two-point function $\langle \mathcal{J}\mathcal{J} \rangle$ in the spinning factorization formula \eqref{eq:spin1fac}. For similar reason, we can define another differential operator $\mathcal{D}_{\mathcal{F}}$ that extracts $\mathcal{F}$, i.e., $\mathcal{F}=\mathcal{D}_{\mathcal{F}}\hat{\mathcal{O}}$,
    \begin{align*}
        \mathcal{D}_{\mathcal{F}} = -\frac{4}{p\sqrt{(p+1)(p-3)}}\left( \partial_{\lambda}^2 \partial_{\pi}^2 - \frac{1}{4(p-2)} \partial_{\alpha\dot{\alpha}} \partial_y \partial^{\dot{\alpha}} \partial^{\alpha} - \frac{1}{16(p-1)(p-2)}\partial^\mu\partial_\mu \partial_y^2 \right),
    \end{align*}
    where $\partial_{\lambda}^2 \equiv \frac{1}{4}\epsilon_{\alpha\beta}\partial^{\alpha}\partial^{\beta}$, and similarly for $\partial^2_\pi$.
    
    Applying the above differentiations at one location in the three- and four-point super-correlators, we can work out the desired component correlators in Mellin space. All the resulting three- and four-point Mellin amplitudes that are needed in the five-point bootstrap computation are presented in the ancillary file on arXiv.

    \section{Gluing SU(2) structures}\label{appx:SU2}
    
    Consider firstly parameterizing a spinor $v$ by 
    \begin{align}
        v = \begin{pmatrix}
          z_1 \\ z_2
        \end{pmatrix}.
    \end{align}
    Our operator $\Op{p}^I$ then becomes a homogeneous polynomial $P$ on $z_1$ and $z_2$ satisfying
    \begin{align}\label{eq:polyrept}
        g\circ P(v) \rightarrow P(g^{-1}v)
    \end{align}
    under the action of $g \in {\rm SU(2)}$. The standard inner product and measure on $\mathbb{C}^2$ 
    \begin{align}
        v^\dagger v = \zb_1 z_1 + \zb_2 z_2, \qquad Dz = \dd z_1 \dd \zb_1 \dd z_2 \dd \zb_2 
    \end{align}
    is invariant under SU(2). We can thus define the inner product of two homogeneous polynomials by
    \begin{align}\label{eq:su2int}
        \int_{v^\dagger v = 1} Dz\  \overline{P(v)}Q(v).
    \end{align}
    This integral serves as an SU(2)-invariant way of gluing SU(2) structures in the sub-amplitudes. One can show that the monomials $z_1^i z_2^{n-i}$ are orthogonal under this inner product
    \begin{align}
        \int_{v^\dagger v = 1} Dz\ \zb_1^i \zb_2^{n-i} z_1^j z_2^{n-j} = \mathcal{N}_0\times i!(n-i)!\, \delta_{ij},
    \end{align}
    where $\mathcal{N}_0$ is an overall normalization factor. Using this fact, the result of the inner product \eqref{eq:su2int} can be easily read off from the coefficients of $z_1^i z_2^{n-i}$
    \begin{align}\label{eq:su2sum}
        \int_{v^\dagger v = 1} Dz\  \overline{p(v)}q(v) =  \mathcal{N}_0 \sum_{i=0}^{n} i!(n-i)!\, \overline{[P]}_i [Q]_i.
    \end{align}
    Here $[P]_i$ denotes the coefficient of $z_1^i z_2^{n-i}$ in $P(v)$. By transforming the conjugate polynomial $\overline{P(v)}$ to be in the same representation as $Q(v)$ (since the representation is pseudo-real), and taking $v = \qty(\substack{1\\ \hat{z}})$ in \eqref{eq:su2sum} we obtain \eqref{eq:polyglue}. The overall normalization can be fixed by considering gluing $(v_1\cdot v)^n$ with $(v\cdot v_2)^n$, which turns out to be $\frac{1}{n!}$.

    \end{appendix}

\end{document}